\documentclass[11pt,twoside]{article}

\usepackage{asp2006}
\usepackage{epsf}
\usepackage{psfig}
\usepackage{lscape}
\usepackage{graphicx}

\newcommand{\brg}{Br$\gamma$}
\newcommand{\kms}{km~s$^{-1}$}
\newcommand{\bra}{Br$\alpha$}
\newcommand{\pfg}{Pf$\gamma$}
\newcommand{\co}{$^{12}$CO}

\begin{document}
\title{An infrared view of (candidate accretion) disks around massive young stars.}   
\author{A. Bik$^1$,  A. Lenorzer$^2$, W.F. Thi$^3$, E. Puga Antol\'in$^4$,  L.B.F.M. Waters$^{5,4}$, L. Kaper$^5$, L. N. Mart\'in Hern\'andez$^2$}   
\affil{$^1$ European Southern Observatory, Karl-Schwarzschild
           Strasse 2, Garching-bei-M\"unchen, D85748, Germany,\\
           $^2$   Instituto de Astrof\'isica de Canarias, 38200 La Laguna, Tenerife, Spain \\
            $^3$  Institute for Astronomy, The University of Edinburgh, Royal Observatory, Blackford Hill, Edinburgh EH9 3HJ, United Kingdom  \\
           $^4$ Instituut voor Sterrenkunde, Celestijnenlaan 200D, B-3001 Leuven, Belgium\\
           $^5$   Astronomical Institute "Anton Pannekoek", University of Amsterdam, Kruislaan 403, 1098 SJ Amsterdam, The Netherlands
            }    

\begin{abstract} 

Near-infrared surveys of high-mass star-forming regions start to shed
light onto their stellar content. A particular class of objects found
in these regions, the so-called massive Young Stellar Objects (YSOs) are
surrounded by dense circumstellar material. Several near- and
mid-infrared diagnostic tools are used to infer the physical
characteristics and geometry of this circumstellar matter.
Near-infrared hydrogen emission lines provide evidence for a
disk-wind. The profiles of the first overtone of the CO band-heads,
originating in the inner 10 AU from the central star, are well fitted
assuming a keplerian rotating disk. The mid-infrared spectral energy
distribution requires the presence of a more extended envelope
containing dust at a temperature of about 200 K. CRIRES observations of CO
fundamental
absorption lines confirm the presence of a cold envelope.
We discuss the evolutionary status of these objects.
\end{abstract}


\section{Introduction}   

The observational study of high-mass star-forming regions is hampered by the fact that they are so unique, and thus, on average distant. Additionally, the actual formation happens deeply obscured inside molecular clouds behind hundreds of magnitudes of visual extinction.  The actual formation sites are only observable at radio and (sub)mm wavelengths. These observations, however, only provide indirect information about the forming stars. The Lyman continuum photon rate derived from the radio flux or the integrated infrared luminosity provides an estimate on the properties of the massive star(s). Only after the stars are formed and start to clear out their environment, they become observable at shorter wavelengths.  The near-infrared window is ideally suited to study the massive stars in this evolutionary phase.
The photosphere and close circumstellar environment of the young massive stars can be detected as the extinction is modest and the thermal emission of the surrounding molecular cloud is not dominant. 

Recent imaging and spectroscopic surveys have revealed the stellar content of high-mass star-forming regions \citep{Hanson02, Kendall03, Alvarez04,Bik04,Blum04, Ostarspec,Brgspec, Kaper07}. Several types of objects were classified in these complex regions. A large number of OB  stars have been identified by their photospheric lines \citep{Watson97,Hanson02,Ostarspec} and their properties do not seem to differ from OB main sequence field stars observed in the optical.  Another class of high-mass objects identified are the massive Young Stellar Objects. 
Their near-infrared colors indicate an infrared excess, and their near-infrared spectra are dominated by emission lines from ionized (hydrogen, helium)  as well as neutral species, e.g. CO \citep{Kendall03,Blum04,Brgspec}. 

These spectroscopic features as well as the infrared excess can be used to derive the physical parameters of the cirumstellar material and try to understand the evolutionary status of these objects. The detection
and study of disks around young massive stars that are remnants of the
formation process is key to our understanding of the formation of
massive stars \citep{Zinnecker07}.

%

\section{Near- and Mid-Infrared diagnostic tools}

In this section the different diagnostic tools available in the near and mid-infrared window will be discussed.
Line ratios of ionized hydrogen lines are used to trace the physical conditions and geometry of the ionized circumstellar material. The line profiles of the  CO first overtone emission provide kinematic information about the dense, neutral gas in the inner 5 - 10 AU. Extended envelopes, located further away from the central star can be detected at longer wavelengths.

\subsection{Ionized gas}

One of the spectral characteristics of the massive YSOs is that they show lines emitted by the recombination of hydrogen and helium. In the K-band, not only \brg, but also weaker lines of the hydrogen Pfund series (Pf23 - Pf 33) are observed, as well as HeI lines in some cases \citep{Brgspec}.
The HeI lines indicate the presence of a hot ionizing source with a spectral type of at least late - mid O \citep{Hanson02}. 

The hydrogen lines can be used to obtain an estimate on the density of the emitting gas. The Pfund lines are high-density tracers which are only emitted in very dense gas (N$_{e}$ up to 10$^{9}$ cm$^{-3}$). These lines likely originate in a circumstellar disk. This is, moreover, supported by the very broad line profiles (300 - 500 \kms) or by double peaked profiles in a few objects.  \brg, on the other hand, can be emitted in a large range of environments with different densities. It can be associated to diffuse HII regions, stellar winds, disk-winds,  or very dense gaseous disks around Be stars. 
 Also here, the observed spectral profile can help to discriminate between possible scenarios.  The observed velocities of \brg\ are in the range of 100 - 200 \kms. This excludes the possibility that the lines are of nebular origin. The recombination lines in an HII regions have a broadening similar to the local sound speed which is on the order of 20 \kms\ for ionized gas. Also, the origin in a stellar wind can be excluded as the velocity profiles of stellar winds show very broad lines (500 - 2000 \kms). 
Double peaked \brg\ emission in some massive YSOs clearly favors the disk-scenario \citep{Blum04}. However, not all the \brg\ lines show this profile and a contribution of the disk-wind cannot be excluded for these. The observed velocity of the \brg\ line is consistent with the disk-wind scenario.

A disk-wind is caused by the interaction between the UV photons and the ionized upper layer of the disk. This results in a radiation driven, outflowing disk-wind with velocities up to 200 \kms \citep{Drew98}.
In a disk-wind, the average density of the emitting ionized gas is lower than
in a geometrically thin circumstellar disk. This will affect the relative strength
of the hydrogen lines.
While in low density regions (e.g. in an HII region), the relative strength of
hydrogen lines  follows the optically thin (case B) approximation \citep{Storey95}; in high density regions,
where the gas is optically thick, the relative strength of the lines is determined by the emitting surface
of the gas.


\begin{figure}
\includegraphics[width=0.9\columnwidth]{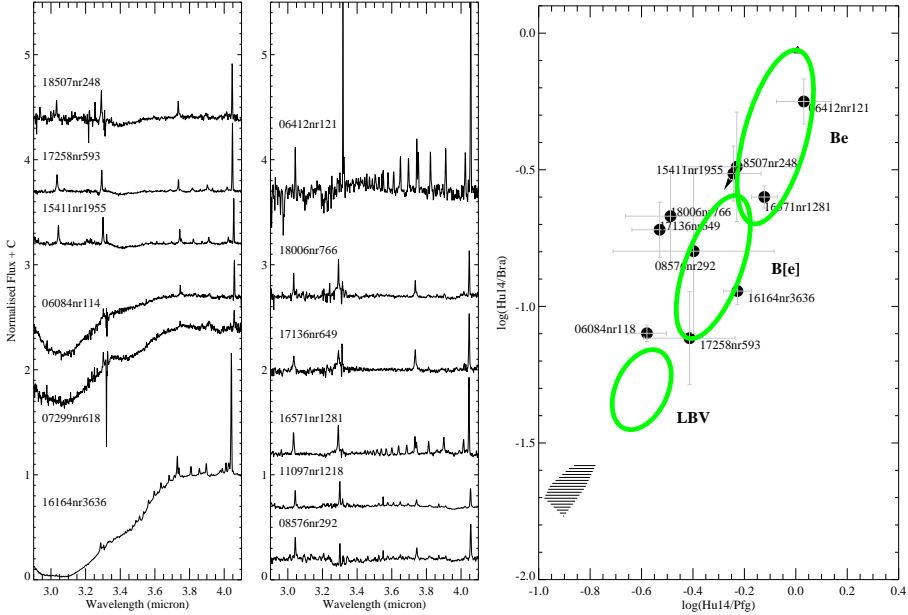}
 \caption{\emph{left 2 pannels:} L-band spectra taken with ISAAC at the VLT of 12 massive YSOs.   The broad absorption are molecular solid-state transition features caused by material in the diffuse ISM or molecular clouds. These absorption features can be used to derive an accurate line-of-sight extinction value. The emission lines are mainly hydrogen emission lines. The lines needed to construct the diagnostic diagramme are \bra (4.05 \micron), \pfg (3.754 \micron) and Hu14 (4.02 \micron).  \emph{right:} The L-band diagramme as described in the text. The gray circles represent the loci of the massive stars with known cirumstellar geometry for which \citet{Lenorzer02} constructed the diagramme: LBVs, B[e] and Be stars. Overplotted are the massive YSOs.}
\end{figure}

\citet{Lenorzer02} constructed a diagnostic diagram using 3 specific hydrogen lines in the L-band (\bra, Hu14 and \pfg) to demonstrate this method. The ratio of Hu14 over \bra\ spans a range of 2 orders of magnitude (from $\sim$ 0.02 in case of an optically thin HII region to unity in the case of an optically thick emitting medium). In order to calibrate the diagram and to relate it to different circumstellar environments, \citet{Lenorzer02} used the observed line ratios of massive stars with a  known circumstellar  geometry (Fig 1b).  Three different types of objects were chosen; Luminous Blue Variables (LBVs), evolved massive stars with a very strong stellar wind, B[e] stars with an equatorially flattened dense stellar wind or disk and the Be stars with an optically thick, geometrically thin ionized disk. The location of these respective objects is plotted in Fig 1b as circles. The Be stars fall as expected close to the optically thick location while the LBV stars are located more in the direction of the optically thin material.

We observed a sample of 12 massive YSOs in the L-band with ISAAC at the VLT (Fig. 1a) and measure the line ratios to compare them with the location of the well studied massive stars in the diagram. 
In Fig 1b the massive YSOs which have all the 3 hydrogen lines in their spectrum are overplotted on the diagram.

The location of the massive YSOs overlaps that of the B[e] and Be
stars. First this confirms that the ionized emitting region of the
massive YSOs is, at least partially, confined into a disk. The span in
location implies large differences from object to object for the
averaged densities of the ionized circumstellar gas.

A possible explanation for this, is that some objects are
surrounded by a geometrically thin and confined disk (similar to Be
disk) while others have disks that are more puffed up as expected when the
disk is under the process of being photo-evaporated.

%

\subsection{CO first overtone emission}

\begin{figure}
\begin{center}
\includegraphics[width=1\columnwidth]{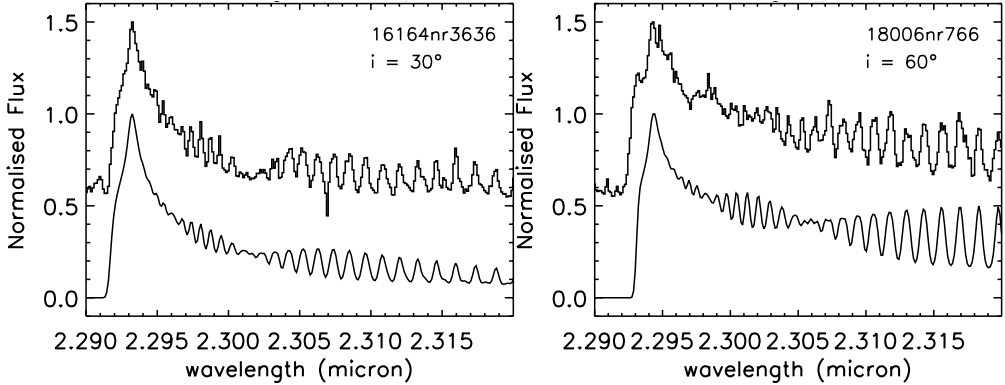}
\end{center}
 \caption{Observed CO bandhead profiles from 2 massive YSOs \citep{Coletter}. The top spectra are the observed spectra, the bottom spectra are the best fitting models. These models indicate that the CO is located in a circumstellar disk within 5 - 10 AU from the central star.}
 \end{figure}

Additional to the ionized lines, lines emitted by neutral species are observed in the spectra of massive YSOs. Among the  most prominent features in the near-infrared are the CO first-overtone emission lines. These bands are observed in about 25\% of all the massive YSOs \citep[e.g][]{Chandler93,Chandler95,Coletter,Blum04}.  The CO bandheads are emitted in neutral material with temperatures between 2000 and 5000 K and densities of about 10$^{10}$ cm$^{-3}$. This hot neutral gas is located in the inner, dust-free regions of the circumstellar environment, relatively close to the star. 

The kinematic information provided by high resolution spectra of the bandheads can be used to constrain the location of the emitting material in the circumstellar matter. Moreover, intrinsically, the blue side of the bandhead is very steep and different velocity profiles give rise to a different shape of the bandhead. Keplerian disks create a blue wing on top of the profile (Fig 2). This blue wing cannot be explained by a gaussian velocity distribution.  \citet{Coletter} and \citet{Blum04} succesfully  fit the high-resolution CO spectra with a emission profile from a rotating disk, assuming optically thin emission. 

This model also shows that the CO  is emitted very close to the central star, in the inner 5-10 AU of the circumstellar environment.  At these distances from the central star, the CO molecules need to be shielded from the UV photons in order not to be photo-dissociated.  Therefore, this gas cannot be located at the surface of the disk, that would be ionized by the stellar radiation.  The optical thickness rises steeply in the radial direction, shielding the midplane of the disk from the UV radiation, allowing the material to become neutral and the CO molecules to survive.

\subsection{Spectral energy distributions}

\begin{figure}
\begin{center}
\includegraphics[width=0.9\columnwidth]{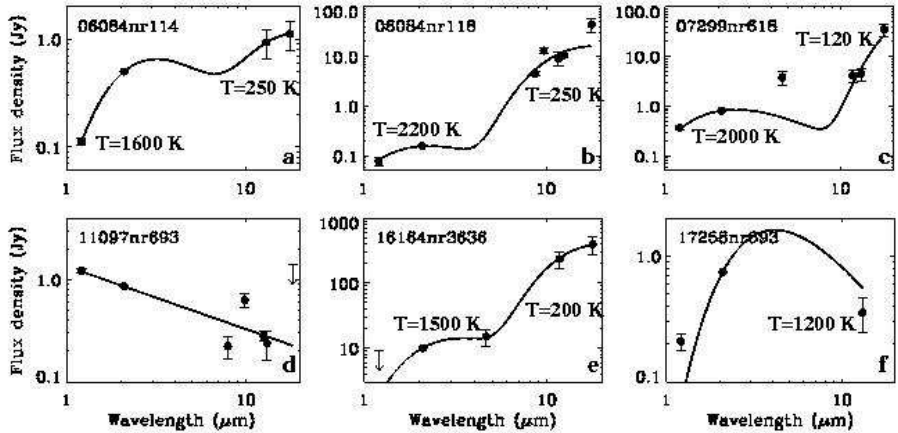}
\end{center}
 \caption{Spectral energy distributions of the candidate massive YSOs. The data points are fit by a simple model. Panels a,b,c and e show a model fit with a 2 temperature black body, the data points of panel d are fitted with a powerlaw and the fit in panel f is a single temperature black body.}
 \end{figure}

One of the main characteristics of massive YSOs is that they have an infrared excess. Their Spectral Energy Distribution  (SED) is not dominated by photospheric emission, but by the emission of circumstellar matter, mainly dust emission.   The near-infrared emission lines, as discussed in the previous subsections trace the ionized or hot neutral gas in the circumstellar environment. The infrared SED traces the warm material (down to a few 100 K) in the cirmstellar environment located further away from the central star. 

We have performed photometry of a sample of massive YSO covering the wavelength range from 1 to 20 \micron\ using the ESO instruments SOFI, TIMM2 and VISIR (see Fig. 3 for some examples).  The observed SEDs are dereddened using extinction values determined by \citet{Brgspec}.  The SEDs show a large range in spectral slopes, suggesting large differences in the properties of the circumstellar environment.  They vary from blue slopes (Fig 3, panel d) to very red slopes (panels b,c and e).

The SED of the bluest object in our sample (11097nr693) cannot be explained by dust emission. The blue slope can only be produced by optically thick free-free emission from e.g. a gaseous disk like in Be stars.   The SED can be fitted with a powerlaw, with a slope typical for Be stars \citep{Waters91}.  Surprisingly, the SED around  10 \micron\ suggests the presence of silicate emission, not expected in a gaseous disk. This could be caused by a contribution of the high background from the surrounding HII region.

The other objects have a much redder SED, and their SED can be explained by dust emission. Overplotted to the observed data in Fig. 3 is the best fit using a one or two temperature blackbody. The objects fitted  with a 2 temperature blackbody show that a  colder component (T  $\sim$ 200 K) tracing dust further away is needed to explain the shape of the SED.  These objects also are spatially resolved at mid-infrared wavelengths and their observed size becomes larger with wavelength.  For one of the reddest objects, 16164nr3636, the size of the 10 and 20 \micron\ emission extends up to 7500 AU away from the central star. Far-infrared observations of this source reveal a strong point source with a derived mass of about 200M\sun\ \citep{Karnik01}. 


\subsection{Absorption spectroscopy with CRIRES}

\begin{figure}
\begin{center}
\includegraphics[width=1\columnwidth]{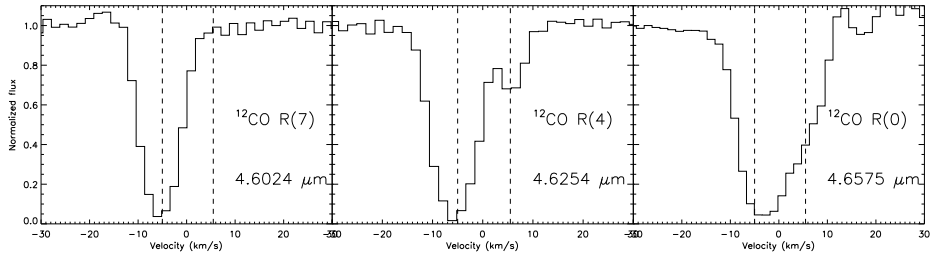}
\end{center}
 \caption{Absorption profiles of 3 \co\ lines observed with CRIRES at high spectral resolution (6 \kms). Two absorption components are detected, the strongest component at +5.5 \kms\ is saturated, the weaker component at -5.5 \kms\ is only detected in the R(4) to the R(0) lines, suggesting a very low temperature.}
 \end{figure}

Another way to study the outer regions of the circumstellar material is the analysis of absorption lines of e.g. CO towards a bright background. As part of the Science Verification observations \citep{Siebenmorgen07} for CRIRES, one of the latest VLT instruments,  we observed one massive YSO (IRAS 16164-5046) at 4.5 \micron\ to detect the \co\ lines of the fundamental transitions. The spectral resolution of these observations is R = 50,000 (6 \kms).  The \co\ lines we  observe are in absorption and show multiple components. In Fig. 4, the R(7), R(4) and R(0) lines of  \co\ lines are plotted.   In all the spectra, a saturated absorption profile is present at -5.5 \kms. In the spectrum of the R(4) line a second absorption is present at +5.5 \kms. This line is not present in the R(7) spectrum but it appears even stronger in the R(0) line. This suggests that this line is absorbed in very cold material of a few 10 K.  The cold envelope detected in the far-infrared observations of \citet{Karnik01} is likely the material that causes the saturated absorption profile of the  \co\ lines. The other component could be related to the source, but also to a cold molecular cloud in the line of sight.

\section{Summary and Evolutionary status}

Several different diagnostic tools are discussed to obtain information about the physical and kinematical properties of the circumstellar material. The different diagnostics trace different physical conditions and therefore different locations in the circumstellar environment of the massive YSOs.
Based on the collection of  observations a sketch can be drawn on how the circumstellar environment of the massive YSOs could look like. In Fig. 5 the different environments and their distance to the central star are sketched. Note that not all the objects have all the characteristics discussed here, some only show a subset of them.


\begin{figure}
\includegraphics[width=1\columnwidth]{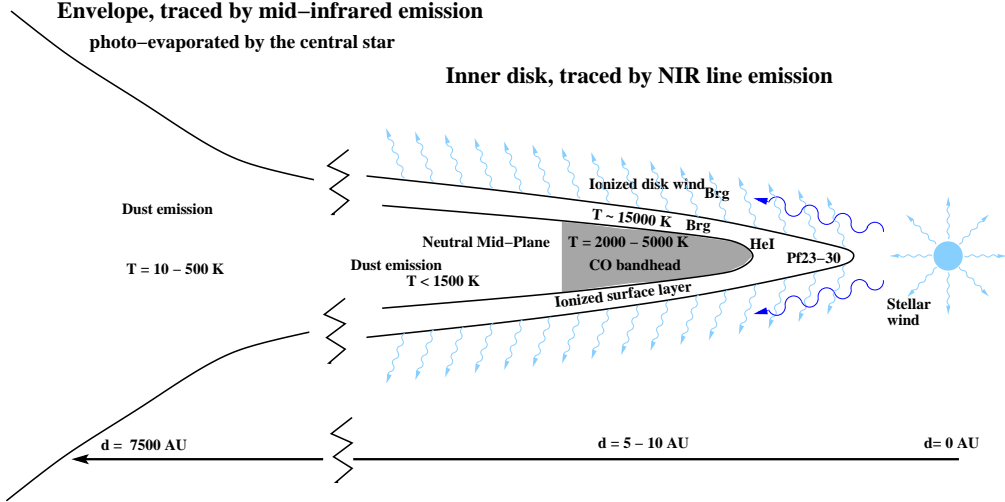}
 \caption{A schematic view of the different line-forming regions in
the circumstellar disk surrounding a massive YSO. The CO is located in
the dense mid plane of the disk, while the hydrogen lines are emitted
by the ionized surface layer and/or disk wind.}\label{fig3}
\end{figure}

Stellar photons ionise the surface layer of the inner disk, which recombination
is observed in the hydrogen and HeI lines; the radiation pressure
exerced onto the surface layer also drives matter away in the form of
a disk wind. In some objects the inner parts of the disk are so dense
that the mid-plane can become neutral as it is shielded for the
UV photons. In these regions (the inner 5 - 10 AU) originate the CO bandhead
emission showing evidence for keplerian rotation. Further
out, the midplane of the disk cools even more and dust is able to survive
giving rise to a red SED observed in the mid-infrared. A few objects
show evidence for a cold dust component suggesting the presence of a
more extended envelope. Most objects, however, are not
detected in the mm regime, suggesting that they lack a large reservoir
of very cold dust or gas.  

This illustrates the difference with the objects discussed by Zhang
(2007) where large disks are detected in the (sub) mm regime. Still in
the process of formation, these objects are too embedded to be
detected in the near-infrared. Likely these large and cold outer
regions of the circumstellar material are (being) photo-evaporated
away.  Photo-evaporation first leads to the disruption of the outer
disk, while the inner disk can survive much longer (Hollenbach et
al. 1994). The inner disk is what is detected at near- and
mid-infrared wavelengths. Objects for which the observations indicate
the presence of an envelope might be the younger objects of our
sample. Another important factor in the photo-evaporation process is
the amount of UV photons emitted by the central star. This can
drastically influence the dispersion time scales. A larger sample
would be needed to disentangle these effects.


\begin{thebibliography}{}
\bibitem[Alvarez et al.(2004)]{Alvarez04} Alvarez C., Feldt, M., Henning, T., Puga, E., Brandner, W., Stecklum, B., 2004, \apjs, 155, 123
\bibitem[Bik,(2004)]{Bik04} Bik, A. 2004, PhD thesis, "The stellar content of high-mass star-forming regions", University of Amsterdam, The Netherlands
\bibitem[Bik \& Thi(2004)]{Coletter} Bik, A. \& Thi, W-.F., 2004, \aap, 427, L13
\bibitem[Bik et al.(2005)]{Ostarspec} Bik, A., Kaper, L., Hanson, M.M., Smits, M., 2005, \aap, 440, 121
\bibitem[Bik et al.(2006)]{Brgspec} Bik, A., Kaper, L. Waters, L.B.F.M., 2006, \aap 455, 561
\bibitem[Blum et al.(2004)]{Blum04} Blum, R.D., Barbosa, C.L., Damineli, A., Conti, P.S., Ridgway, S., 2004, 617, 1167
\bibitem[Chandler et al.(1993)]{Chandler93} Chandler, C.J. Carlstrom, J.E., Scoville, N.Z., Dent, W.R.F, Geballe, T.R., 1993, ApJL, 412, L71
\bibitem[Chandler et al.(1995)]{Chandler95} Chandler, C.J. Carlstrom, J.E., Scoville, N.Z., 1995, \apj, 446, 793
\bibitem[Drew et al.(1998)]{Drew98} Drew, J. E., Proga, D., Stone, J. M., 1998, \mnras, 296,L6 
\bibitem[Hanson et al.(2002)]{Hanson02} Hanson, M. M., Luhman, K. L., Rieke, G. H., 2002, \apjs, 138, 35
\bibitem[Hollenbach et al.(1994)]{Hollenbach94} Hollenbach, D., Johnstone, D., Lizano, S., Shu, F., 1994,\apj,428,654
\bibitem[Karnik et al.(2001)]{Karnik01} Karnik, A. D., Ghosh, S.K., Rengaranjan T.N., Verma, R., P., 2001, \mnras, 326,293
\bibitem[Kaper et al.(2007)]{Kaper07} Kaper, L. Bik, A. Hanson, M.M., Comer\'on, F., 2007, \aap, in press
\bibitem[Kendall et al.(2003)]{Kendall03} Kendall, T. R., de Wit, W. J., Yun, J. L., 2003, \aap 408, 313
\bibitem[Lenorzer et al.(2002)]{Lenorzer02} Lenorzer, A., de Koter, A., Waters, L. B. F. M., 2002, \aap, 386, L5
\bibitem[Siebenmorgen et al.(2007)]{Siebenmorgen07} Siebenmorgen. Ralf., Smette, Alain; K\"aufl, Hans Ulrich, Seifahrt, Andreas, Uttenthaler, Stefan, Bik, Arjan, Casali, Mark, Hubrig, Swetlana, Jung, Yves, Kerber, Florian et al., 2007, The Messenger, 128, 17
\bibitem[Storey \& Hummer(1995)]{Storey95} Storey, P.J., Hummer, D.G., 1995, \mnras 272, 41
\bibitem[Waters et al.(1991)]{Waters91} Waters, L.B.F.M, Marlborough, J.M., van der Veen, W.E.C., Taylor, A.R., Dougherty, S.M., 1991, \aap, 244, 120
\bibitem[Watson \& Hanson(1997)]{Watson97} Watson, A.M., Hanson, M.M., 1997, ApJL, 490, L165
\bibitem[Zhang (2007)]{Zhang07} Zhang, Q, 2007, These proceedings
\bibitem[Zinnecker \& Yorke(2007)]{Zinnecker07} Zinnecker, H, Yorke, H.W., 2007, \araa, 45, 481 




\end{thebibliography}
\end{document}